\documentclass[11pt]{article}
\usepackage{amsmath, amssymb, graphicx, natbib, setspace, multirow, xcolor}
\usepackage[margin=1in]{geometry}
\usepackage{amsmath, amssymb}
\usepackage{algorithm}
\usepackage{algpseudocode}
\usepackage{xcolor}

\title{$p$-PSO: A Penalized Particle Swarm Optimization Technique for Finding D-Optimal Designs with Mixed Factors in Generalized Linear Models}

\author{Shrabanti Chowdhury\\Icahn School of Medicine at Mount Sinai\\New York, NY 10029-5674 \and Abhyuday Mandal\\University of Georgia\\Athens, GA 30602-7952}

\date{}

\begin{document}

\maketitle

\begin{abstract}
Finding D-optimal designs for generalized linear models (GLMs) is challenging due to the dependence of the Fisher information matrix on unknown parameters and the lack of closed-form solutions, particularly when input factors include both discrete and continuous variables. Although classical algorithms and recent metaheuristic approaches have offered partial solutions, there remains a need for robust and computationally efficient methods. In this paper, we propose a penalized Particle Swarm Optimization (PSO) approach, named $p$-PSO. Here we introduce a new, general-purpose penalty formulation for constrained optimization and demonstrate its effectiveness in optimal design problems. The formulation is algorithm-agnostic and applicable to a broad class of black-box optimization methods. Results show that the method is highly efficient, with its primary contribution being a penalty formulation that enables the direct use of an off-the-shelf PSO algorithm and extends naturally to more general constrained optimization tasks.
\end{abstract}


 \vspace{-5mm} 

\section{Introduction}

Generalized linear models (GLMs) provide a flexible framework for modeling mean responses of both discrete and continuous random variables, with particular emphasis on categorical outcomes. Since inference on the model parameters strongly depends on the experimental design, the choice of design is crucial. While methods for analyzing data from GLMs are well developed \citep{mccullagh1989, agresti2002, lindsey1997, mcculloch2001, dobson2008, myers2002}, results on optimal design for GLMs remained limited until early twentieth century \citep{khuri2006, woods2006, atkinson2007}. Since then, there has been a surge of research in this area. Theoretical contributions include complete class results and characterizations of optimal designs \citep{stufken2012,  yang2012, yang2016, yang2017d, bu2020d}. On the computational side, several powerful algorithms have been developed, such as $d$-QPSO algorithm \citep{lukemire2020}, and more recently the ForLion algorithm \citep{huang2024}. While these algorithms are powerful, they are relatively complex to implement.  

In contrast, the goal of this paper is to present a very simple approach that is easy to code and apply in practice. Specifically, we propose a penalized particle swarm optimization ($p$-PSO) method that augments a standard off-the-shelf PSO algorithm with a penalty function to handle design constraints. Although the idea is straightforward, it provides a general framework that can be applied beyond the design problems considered here, offering a practical route for constrained optimization in experimental design and related fields.

Design of experiments plays a central role in scientific investigation, providing efficient strategies to collect data for accurate estimation of model parameters. In generalized linear models (GLMs) with binary or ordinal outcomes, finding D-optimal designs is particularly difficult when both discrete and continuous factors are present. Classical results exist for settings where all factors are discrete or all continuous (e.g., \citealp{yang2012, yang2016, yang2017d}), but identifying mixed-factor designs still remains a challenging problem. It is extremely difficult to obtain general theoretical results, and state-of-the-art efficient computational algorithms  \citep{lukemire2020, huang2024} are very complex for practitioners to use.

In the context of identifying optimal designs for GLMs, a key challenge is that the Fisher information matrix depends on unknown parameters, requiring practitioners to rely on educated guesses on the model parameters that lead to locally optimal designs \citep{chernoff1953}, or more computationally burdensome Bayesian designs \citep{chaloner1995}. Traditional numerical approaches, such as Fedorov--Wynn exchange \citep{fedorov1972}, multiplicative algorithms \citep{titterington1976}, and the cocktail algorithm \citep{yu2011}, are often computationally demanding and limited in scope. More recently, nature-inspired algorithms such as particle swarm optimization (PSO; \citealp{kennedy1995}) and its variants have been proposed for design generation \citep{chen2023particle}. In particular, quantum-behaved PSO (QPSO; \citealp{ sun2004quantum2,sun2004b_global, sun2012quantum1}) has inspired recent work on mixed-factor designs with binary responses \citep{lukemire2019} and ordinal outcomes \citep{lukemire2020}.

In this work, we introduce $p$-PSO, a penalized particle swarm optimization method, for finding D-optimal designs for GLMs with mixed factors. The method extends standard PSO by incorporating penalty terms into the fitness function, ensuring feasibility of designs while maintaining robust exploration of the design space. Note that the introduction of the penalty function allows us to turn the combinatorial problem into a continuous optimization problem. We show that $p$-PSO can efficiently identify both approximate and exact designs.

\section{Preliminaries and Notation}

\subsection{Exact versus Approximate Designs}

In optimal design theory, one distinguishes between \emph{exact designs} and \emph{approximate designs}. 
An exact design specifies integer allocations of experimental runs to a finite set of design points. 
Suppose there are $m$ candidate support points $\boldsymbol{x_1},\dots,\boldsymbol{x_m}$, and let $n_i$ denote the number of 
replications at $\boldsymbol{x_i}$, with the total sample size fixed at $N = \sum_{i=1}^m n_i$. Finding optimal 
exact designs is often challenging because the allocation variables $n_i$ are restricted to integers, 
which makes the optimization problem combinatorial in nature. 

To overcome this difficulty, \citet{kiefer1959} introduced the framework of \emph{approximate designs}, 
where one works instead with proportions $p_i = n_i/N$ that satisfy $\sum_{i=1}^m p_i = 1$. 
In this formulation, an approximate design is represented as
\[
\xi = 
\left\{ 
\begin{array}{cccc}
	\boldsymbol{x}_1 & \boldsymbol{x}_2 & \cdots & \boldsymbol{x}_m \\
	p_1 & p_2 & \cdots & p_m
\end{array} 
\right\},
\]
where $p_i$ is the fraction of the total experimental effort allocated to design point $\boldsymbol{x}_i$. 
The key advantage of this representation is that the Fisher information matrix becomes a convex 
combination of the information matrices at the support points:
\begin{equation}
	I(\xi, \theta) = \sum_{i=1}^m p_i I(\boldsymbol{x_i}, \theta), 
	\label{eq:info}
\end{equation}
where 
\begin{equation}
	I(\boldsymbol{x_i}, \theta) = 
	\frac{\partial f(\boldsymbol{x_i}, \theta)}{\partial \theta} 
	\frac{\partial f(\boldsymbol{x_i}, \theta)}{\partial \theta^{\!T}}
	\label{eq:info-point}
\end{equation}
is the information matrix from a single observation at $\boldsymbol{x_i}$. 
This formulation allows one to exploit tools from convex analysis and calculus, 
leading to a unified theory of approximate designs 
\citep{kiefer1974, fedorov1972, silvey1980, pukelsheim2006, atkinson2007}. Note that in this paper we need the derivation of the information matrix under GLM, see Section~\ref{Sec:5} for details (cf: equation~(\ref{eq.5.1.FI})).

Kiefer’s general equivalence theory provides a broad class of optimality criteria indexed by 
a parameter $p$:
\begin{equation}
	\Phi_p(I(\xi,\theta)) = 
	\left( \frac{1}{k} \, \text{Tr} \!\big(I^{-p}(\xi,\theta)\big) \right)^{1/p}, 
	\quad 0 \le p \le \infty,
	\label{eq:phip}
\end{equation}
where $k$ is the number of model parameters. 
Different values of $p$ correspond to classical criteria: $p=1$ yields 
A-optimality, minimizing the average variance of the parameter estimates; 
$p\to 0$ gives D-optimality, maximizing $|I(\xi,\theta)|$ and minimizing the volume 
of the joint confidence ellipsoid; and $p \to \infty$ leads to E-optimality, 
minimizing the largest eigenvalue of the covariance matrix. 
Thus, approximate design theory provides a unifying framework in which many optimality 
criteria can be expressed and analyzed.

\subsection{Locally Optimal Designs}

For nonlinear models, the Fisher information matrix depends on the unknown parameter vector $\theta$, 
which complicates the design problem. A common strategy is to assume nominal values of $\theta$, 
obtained for example from previous studies or expert knowledge, and then optimize the design 
criterion conditional on these values. The resulting designs are called \emph{locally optimal} 
\citep{chernoff1953}. While locally optimal designs are simple to compute and often highly efficient 
when the assumed values are accurate, they can be sensitive to misspecification of $\theta$. 
In practice, this sensitivity motivates alternative approaches such as Bayesian or maximin designs, 
which aim to provide robustness by averaging or optimizing over a range of plausible parameter values 
\citep{chaloner1995, pronzato1988, imhof2001}. 

Despite these limitations, locally optimal designs remain a cornerstone of the approximate design 
literature, providing valuable insights and tractable solutions in many applications. 
Comprehensive treatments of exact and approximate optimal designs, along with equivalence theory 
and computational algorithms, can be found in standard references such as 
\citet{pukelsheim2006} and \citet{atkinson2007}.

\subsection{Bayesian and Pseudo-Bayesian Designs}

When parameter uncertainty exists, Bayesian and pseudo-Bayesian designs offer alternatives. Bayesian designs maximize expected information with respect to a prior distribution \citep{chaloner1995}, but require expensive integration. Pseudo-Bayesian designs \citep{woods2006} approximate this expectation using representative parameter sets, offering a computationally feasible alternative. These methods are attractive for robustness but can be computationally intensive when combined with large search spaces.

\subsection{Equivalence Theorems}\label{Sec2.4:eq}

An important tool in approximate design theory is the \emph{equivalence theorem}, which provides a way to verify the optimality of a candidate design. 
For the case of D-optimality, the criterion is to maximize the determinant of the Fisher information matrix, or equivalently to minimize 
\[
\Phi_D(\xi) = - \log |I(\xi)|,
\]
where $I(\xi)$ is the information matrix of design $\xi$. 
Consider a regression model with mean function $f(\boldsymbol{x},\theta) = g(\boldsymbol{x})^{T}\theta$, where $g(\boldsymbol{x})$ is a vector of $k$ linearly independent regression functions. 
The equivalence theorem states that a design $\xi_D$ is D-optimal if and only if 
\begin{equation}
	s(\boldsymbol{x}, \xi_D) = g(\boldsymbol{x})^{T} I^{-1}(\xi_D) g(\boldsymbol{x}) - k \leq 0, \quad \forall \boldsymbol{x} \in \mathcal{X},
	\label{eq:equiv}
\end{equation}
with equality holding at the support points of $\xi_D$. 
This result provides a necessary and sufficient condition for optimality: a design is D-optimal precisely when its sensitivity function $s(\boldsymbol{x},\xi_D)$ does not exceed zero over the entire design space and equals zero at the support points.

Equivalence theorems are not limited to D-optimality; analogous results exist for many convex criteria, forming the backbone of modern optimal design theory 
\citep{kiefer1974, fedorov1972, silvey1980, pukelsheim2006, atkinson2007}. 
Although the equivalence theorem is a powerful result for verifying optimality, it does not provide guidance on how to \emph{find} optimal designs in the first place. 
In fact, theoretical characterizations of optimal designs are only available in a handful of special cases. 
As models grow more complex or high-dimensional, analytical solutions become intractable. 
For this reason, the next section turns to numerical and heuristic techniques that can be used to construct optimal designs in practice.

\section{Algorithmic Approaches to Optimal Design}
The design of experiments under GLMs has been the subject of a rich body of research. We review several strands of the literature here, focusing on optimization algorithms used for generating optimal designs, their advantages, and their limitations. Interested readers may refer to \cite{liski2002topics} and \cite{mandal2015optimal} for a detailed overview.

\subsection{Classical Optimization Techniques}
Early approaches to optimal design include the exchange algorithm of \citet{fedorov1972} and Wynn’s algorithm, which iteratively replace design points with alternatives that improve the optimality criterion. These algorithms guarantee convergence to a local optimum, but they can be slow and sensitive to starting conditions when the design space is large. Multiplicative algorithms \citep{titterington1976} use proportional updates of design weights and are computationally efficient, but they may converge prematurely in nonlinear settings. The cocktail algorithm of \citet{yu2011} combines the strengths of exchange and multiplicative algorithms. While it often performs better in practice, its applications remain limited when continuous factors are present because the search space must still be discretized.

\subsection{Gradient-Based and Deterministic Methods}
Gradient-based methods, such as quasi-Newton optimization and sequential quadratic programming \citep{nocedal1999}, have been applied to approximate design problems. These methods are powerful for continuous optimization but require differentiability and can be trapped in local optima, especially in high-dimensional or mixed discrete-continuous design problems. Deterministic global optimization methods such as branch-and-bound or cutting-plane algorithms \citep{floudas1999} provide guarantees but are often infeasible for large experimental design problems.

\subsection{Metaheuristic Approaches}
Metaheuristic algorithms inspired by natural processes have become increasingly popular due to their flexibility. Among these:

\begin{itemize}
  \item \textbf{Simulated Annealing (SA)}: Mimics the physical annealing process, allowing occasional uphill moves to escape local optima \citep{kirkpatrick1983}. SA is simple and versatile but often slow to converge, requiring careful tuning of cooling schedules. One of the earliest notable application of simulated annealing to construct exact D-optimal designs, showing effectiveness in finite and continuous design spaces, was by \cite{meyer1988N}. More recently, \cite{joseph2008Ying} used a simulated annealing algorithm to construct space-filling Latin Hypercube Designs (LHDs) with good projection properties. \cite{joseph2015Ba} proposed Maximum Projection (MaxPro) designs, generated via a stochastic optimization approach (including simulated annealing), for constructing space-filling designs with excellent projection properties.

  \item \textbf{Genetic Algorithms (GA)}: Inspired by evolutionary processes, GA employs selection, crossover, and mutation to evolve candidate solutions \citep{holland1975}. GAs are effective in exploring large, complex spaces but can suffer from premature convergence without diversity-preserving strategies \citep{goldberg1989}. This has been extensively used in design literature. \cite{lin2014GAreview} provides is a comprehensive review that examines how genetic algorithms (GAs) have been applied in statistical experimental design, highlighting when they offer competitive advantages in finding optimal designs and addressing challenges like convergence and computational efficiency. Motivated by GAs, \cite{mandal2006} proposed SELC, a variant of genetic algorithms, for screening high-dimensional design spaces in drug discovery and related experiments by sequentially eliminating unpromising factor level combinations. This was later improved to $G$-SELC algorithm, by integrating prior information (like clustering or known good regions) into the genetic algorithm framework, with Gaussian Process modeling, in order to improve efficiency for searching complex design spaces \citep{mandal2009}. 

  \item \textbf{Differential Evolution (DE)}: While genetic algorithms (GAs) have been widely used and are particularly effective for problems with discrete factors, \citet{storn1997} introduced the idea of Differential Evolution (DE) into the design literature, where the notions of crossover and mutation in GAs are generalized. Differential Evolution works by using vector differences to generate new candidate solutions. Since then, many extensions and applications have been proposed to enhance its performance across different contexts \citep{mitchell1998, feoktistov2006, plagianakos2008}.  Current optimization literature suggests that
  in many situations, DE produces better, more stable solutions than Genetic Algorithms \citep{DE2007}. In the design of experiments, DE has been applied to logistic regression design problems, demonstrating robustness and flexibility. However, DE is sensitive to parameter choices and may struggle when dealing with discrete factors.
  
  \item \textbf{Tabu Search}: Introduced by \citet{glover1989}, tabu search explores neighborhoods while preventing cycles through memory structures. It performs well in combinatorial settings but requires problem-specific tuning.
  
  \item \textbf{Ant Colony Optimization (ACO)}: Models pheromone trails of ants to guide search \citep{dorigo1996ant,DorDicGam99:al,dorigo2004ant}. ACO excels in discrete combinatorial design problems but scales poorly with continuous factors.
  
  \item \textbf{Particle Swarm Optimization (PSO)}: Models the social behavior of swarms to explore the design space \citep{kennedy1995}. It is another population-based metaheuristic inspired by the collective behavior of a flock of birds, where candidate solutions (``particles'') move through the search space by sharing information about their best positions. It is simple to implement, requires few parameters, and is effective for continuous optimization problems. PSO is computationally efficient, requires few assumptions, and naturally accommodates continuous factors. However, it can stagnate near local optima without modification. In the design literature, PSO was popularized by Dr. Weng-Kee Wong \citep{chen2022}. See \cite{mandal2015optimal} for a detailed overview of PSO. Among others, \cite{garcia2019} provides a systematic comparison of several general-purpose optimization algorithms (including PSO, GA, SA, DE, etc.) for computing optimal approximate experimental designs. The study highlights relative strengths, weaknesses, and efficiency trade-offs across different problem settings. 

  \item \textbf{Quantum-Behaved PSO (QPSO)}: A variant introduced by \citet{sun2004quantum2,sun2004b_global} that replaces velocity updates with probabilistic position updates, improving convergence and exploration. \citet{lukemire2019} adapted QPSO to mixed-factor binary response designs ($d$-QPSO), showing significant efficiency gains.
\end{itemize}

\subsection{Advantages and Disadvantages}
Classical algorithms are deterministic and interpretable but struggle with scalability. Gradient-based methods are efficient but require differentiability and good initialization. Metaheuristics like GA, SA, DE, ACO, and tabu search provide global search capabilities but can be computationally demanding. PSO and QPSO strike a balance between exploration and efficiency but need adaptations to handle mixed discrete-continuous factors and constraints. Deterministic global optimization methods guarantee convergence but often do not scale.

This landscape motivates the development of $p$-PSO, a new framework for constrained optimization, demonstrated here using PSO. We propose a general-purpose, algorithm-agnostic penalty formulation and evaluate it on optimal design problems. The method is applicable to a wide range of black-box optimization algorithms. $p$-PSO algorithm enables the direct use of standard PSO implementations, while naturally extending to more general constrained optimization settings. By incorporating penalty functions into the PSO framework, $p$-PSO preserves PSO’s flexibility and improves convergence to feasible solutions.

\section{Proposed Methodology: The $p$-PSO Algorithm}\label{Sec4:pPSO}

A motivating example for our work comes from the \emph{car refueling experiment}, originally described in \citet{grimshaw2001}. The goal of this study was to evaluate the performance of a vision-based automated refueling system for cars. Specifically, the investigators wished to determine whether a computer-controlled nozzle could consistently insert itself into the gas pipe of a vehicle without human intervention. This setup naturally leads to a binary outcome: success if the nozzle correctly inserted, and failure otherwise. The experiment involves ten experimental factors. Four of these are categorical (discrete) variables such as ring type, lighting condition, sharpening, and smoothing of the camera images. The remaining six are continuous variables: reflective ring thickness, lighting angle, gas-cap angle (Z-axis), gas-cap angle skew (Y-axis), car distance, and threshold step value. Together, these factors form a challenging design space that mixes discrete and continuous factors. The ten factors can be summarized as follows. The categorical factors include: 
\begin{enumerate}
	\item {Ring type:} white paper or reflective ring, \\[-20pt]
	\item {Lighting:} standard room lighting versus room lighting with additional flood lights,\\[-20pt]
	\item {Sharpen:} no sharpening versus sharpening of image,\\[-20pt]
	\item {Smooth:} no smoothing versus smoothing of image.
\end{enumerate}

\noindent On the other hand, the continuous factors vary over specified ranges:
\begin{enumerate}
	\item {Lighting angle:} 50 to 90 degrees,\\[-20pt]
	\item {Gas-cap angle (Z-axis):} 30 to 55 degrees,\\[-20pt]
	\item {Gas-cap skew (Y-axis):} 0 to 10 degrees,\\[-20pt]
	\item {Car distance:} 18 to 48 inches,\\[-20pt]
	\item {Reflective ring thickness:} 0.125 to 0.425 inches,\\[-20pt]
	\item {Threshold step value:} 5 to 15.
\end{enumerate}

In this car refueling experiment, the binary response indicates whether the nozzle successfully entered the gas pipe. We model the mean response $\mu$ using a logistic regression framework:
\begin{align*}
	\text{logit}(\mu) &= \beta_0 + \beta_1 \text{(Ring type)} + \beta_2 \text{(Lighting)} + \beta_3 \text{(Sharpen)} + \beta_4 \text{(Smooth)} \\
	&\quad + \beta_5 \text{(Lighting angle)} + \beta_6 \text{(Z-axis angle)} + \beta_7 \text{(Y-axis skew)} \\
	&\quad + \beta_8 \text{(Car distance)} + \beta_9 \text{(Ring thickness)} + \beta_{10} \text{(Threshold step value)}.
\end{align*}

Our aim is to identify $D$-optimal designs for estimating the regression coefficients. The $D$-optimality criterion seeks to maximize the determinant of the Fisher information matrix, equivalently maximizing the information gained about the parameters. This mixture of categorical and continuous variables exemplifies the type of experimental design problems that are computationally difficult to handle. We focus on locally optimal designs, and the nominal parameter values in the fitted logistic regression model were taken to be
\[
\boldsymbol{\beta} = (3, 0.5, 0.75, 1.25, 0.8, 0.5, 0.8, -0.4, -1.00, 2.65, 0.65)^{T}.
\]

Search algorithms such as Particle Swarm Optimization (PSO) are often effective for identifying optimal designs when all factors are continuous. However, when categorical factors are included, the optimization problem becomes significantly harder. The main difficulty is that categorical factors introduce discontinuities into the design space, making it more challenging to explore efficiently. Recent methods such as $d$-QPSO \citep{lukemire2019} have been developed to accommodate both continuous and categorical factors. These approaches often involve specialized update rules for discrete factors, in addition to the standard updates for continuous ones. As mentioned before, while they can be effective, they are also considerably more complex to implement.

Although standard PSO is fast and effective for small-scale problems, it may struggle when the design problem involves many mixed factors. To address such challenges, \citet{sun2004quantum2, sun2004b_global} introduced Quantum-behaved PSO (QPSO), where particles are modeled as probabilistic entities that may occupy any point in the search space, with higher likelihood near their current positions. \citet{lukemire2019} extended this to $d$-QPSO to address complex design problems with both discrete and continuous variables. In contrast to these algorithmic modifications, our work proposes a much simpler approach: we modify the \emph{objective function} rather than the search algorithm itself. This allows standard, off-the-shelf PSO implementations to be used for design problems with discrete factors.

\subsection{Penalized PSO ($p$-PSO)}

The central idea of our penalized PSO ($p$-PSO) method is to transform discrete optimization problems into continuous ones by introducing a penalty function. Discrete factor levels are projected onto the continuous line, allowing PSO to move freely through the space. The objective function is then penalized depending on the distance between the proposed (possibly invalid) value and the nearest valid level.

\begin{figure}[ht]
	\centering
	\includegraphics[width=0.6\textwidth]{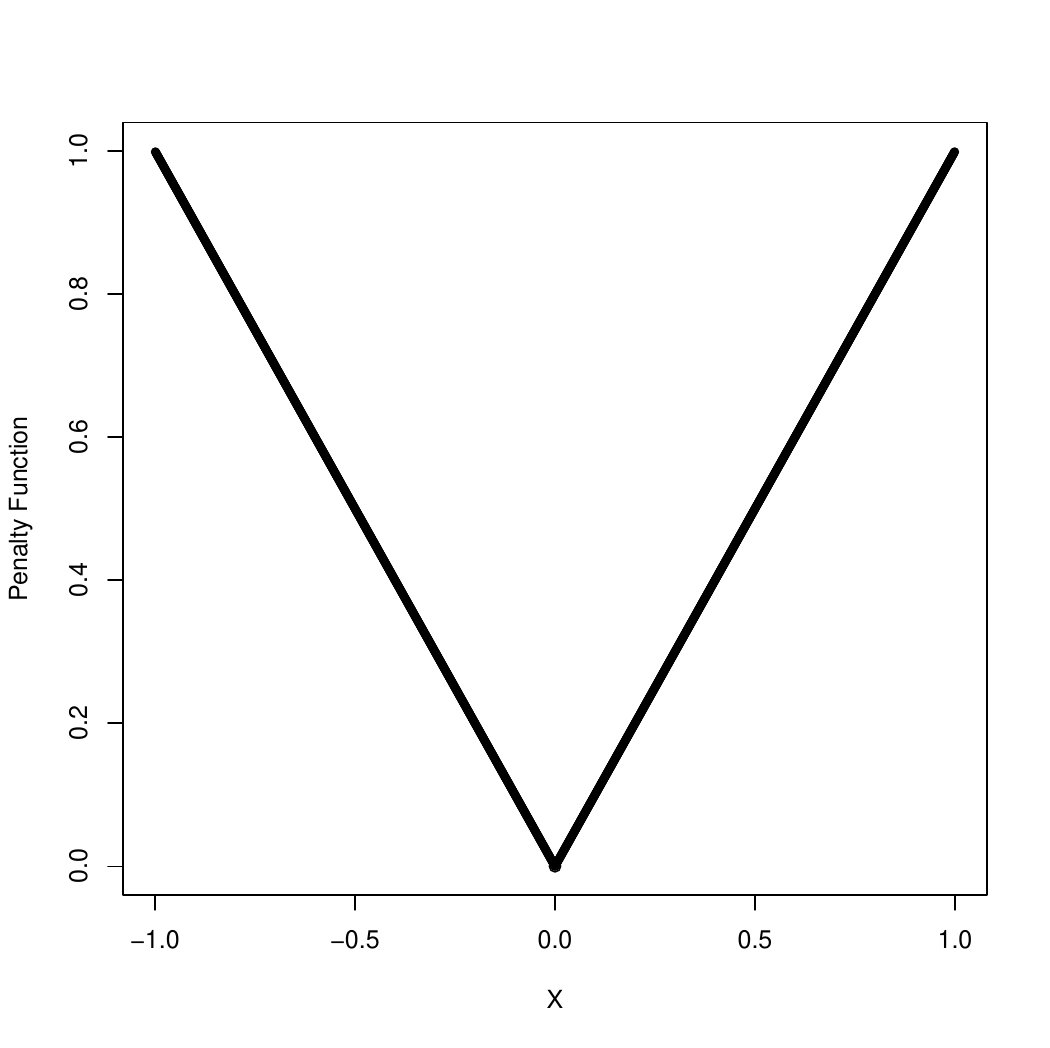}
	\caption{Illustration of the penalty function used in equation~(\ref{Eq:lambda}). It takes value $+1$ for $x=\pm 1$ and is equal to 0 at $x=0$, the furthest point from the feasible ones, $\pm 1$. Also note that $\lambda$ is multiplicative with $g$ in equation~(\ref{eq:g}).}
	\label{fig:penalty}
\end{figure}

Formally, if $g(X)$ denotes the objective function, we replace it with a penalized version:
\begin{eqnarray}\label{eq:g}
g^{*}(X) = g(\tilde{X}) \, \lambda(X),
\end{eqnarray}
where $\tilde{X}$ replaces any settings for invalid factors in $X$ by the nearest valid level, and $\lambda(X)$ is a penalty term with $0 < \lambda(X) < 1$. For example, a possible choice of $\lambda$ is 
\[
\lambda(X) = 1 - \sqrt{(X - \tilde{X})^2}.
\]
For illustration, consider a simple example where $X$ takes only two values $\pm 1$. In this case, we have the following
\begin{eqnarray}\label{Eq:lambda}
	\lambda(x) = \left\{\begin{array}{lcc} 1 - \sqrt{(x+1)^2} & \mbox{if} & x \le 0 \\ 1 - \sqrt{(x-1)^2} & \mbox{if} & x > 0 \end{array} \right.
\end{eqnarray}
This penalization ensures that invalid designs are never superior to valid ones and encourages the search to converge towards admissible solutions.
The penalty function plays a central role in discouraging infeasible solutions. Figure~\ref{fig:penalty} shows its behavior across different values of $x$ as given in equation~(\ref{Eq:lambda}).

Extending this to multiple categorical factors, the penalization can be applied multiplicatively across all $m$ discrete factors:
\[
\phi(X) = g(\tilde{X}) \prod_{k=1}^{m} \left(1 - \sqrt{(X_k - \tilde{X}_k)^2}\right).
\]
where $m=4$ is the total number of discrete factors. The penalized objective function is then
\[
g^{*}(X) = \prod_{i=1}^N \phi(X_i) = \prod_{i=1}^N \left\{ g(\tilde{X_i})\prod_{k=1}^{m}(1 - \sqrt{(X_{i,k} - \tilde{X}_{i,k})^2})\right\},
\]
where $X_1, \ldots, X_N$ are the candidate design points.  Note that if the search algorithm gives an ``invalid design”, such as a design with a categorical setting at $-0.75$, we are guaranteed to find a better design by replacing the invalid value with the closest valid value, since $(\tilde{X})$ was used in the evaluation of $g$. Although we consider two-level factors, this approach is not limited to factors with only two levels. The key point to note is that this framework effectively transforms the design optimization problem into a continuous one, while guaranteeing that the best solutions correspond to valid designs. Thus, off-the-shelf optimization techniques for continuous optimization can be readily used. Moreover, the approach is not limited to two-level factors as illustrated above in equation~(\ref{Eq:lambda}), but can be generalized to categorical factors with more than two levels.

\subsection*{Proposed Methodology: The $p$-PSO Algorithm}

The penalized PSO modifies the standard PSO framework by redefining the fitness function. Let $I_\xi$ denote the Fisher information matrix of design $\xi$, and $\Phi(\xi) = \log \det(I_\xi)$ the D-optimality criterion. The penalized fitness is defined as:
\begin{equation}
F(\xi) = \Phi(\xi) - \lambda P(\xi),
\end{equation}
where $P(\xi)$ is a penalty function capturing violations of feasibility (e.g., negative weights, weights not summing to 1, factor levels outside admissible ranges), and $\lambda > 0$ is a tuning constant.

Particles represent candidate designs, with positions corresponding to support points and weights. At each iteration, positions are updated toward the personal best ($pbest$) and global best ($gbest$) while evaluating the penalized fitness. The penalty function ensures that infeasible designs receive reduced fitness, discouraging exploration of invalid regions.

The algorithm proceeds as follows:

\begin{enumerate}  
\item For each particle $i=1,2,...,N:$
\begin{itemize}
    \item Initialize position $x_i$ randomly in search space
    \item Initialize velocity $v_i$ randomly
    \item Project $x_i$ to nearest valid design $\tilde{x}_i$:
           replace each discrete factor by its nearest valid level
    \item Compute penalized fitness: $F(x_i) = \Phi(\tilde{x}_i) - \lambda P(x_i)$
    \item Determine global best \textit{gbest} among all particles
\end{itemize}

 \item For each particle $i$
\begin{itemize}
    \item Update velocity:
    $v_i \xleftarrow{} wv_i + r_1(pbest - x_i) + r_2(gbest - x_i)$, $r_1, r_2 \sim U(0, 2)$ 
    \item Update position: $x_i \xleftarrow{} x_i + v_i$
    \item Project $x_i$ to nearest valid design $\tilde{x}_i$
    \item Compute penalized fitness: $F(x_i) = \Phi(\tilde{x}_i) - \lambda P(x_i)$
    \item Update personal best: $F(x_i) < F(pbest_i)$, then $pbest_i = x_i$ 
    \item Update global best: $F(pbest_i) < F(gbest)$, then $gbest = pbest_i$
\end{itemize} 

\item After $T$ iterations, return the best feasible solution $gbest$.
\end{enumerate}

A detailed pseudocode description of the algorithm can be found in the appendix.

\section{Applications}\label{Sec:5}
We evaluate $p$-PSO on case studies drawn from experimental design literature and our motivating examples.

\subsection{Odor Removal Experiment}
The odor removal study \citep{wang2016} investigated effects of algae type, scavenger, resin, compatibilizer (discrete factors), and storage temperature (continuous factor) on the success of odor elimination in bio-plastics. The outcome was binary, whether or not the odor was removed successfully. In the absence of standard literature for such mixed factor optimal designs, scientists fixed the temperature at 25$^\circ$C in order to use a standard two-level factorial design. Using information from their study, we take nominal values $\beta=(-1,2,0.5,-1,-0.25,0.13)^T$ in order to apply the $p$-PSO algorithm for obtaining a locally D-optimal approximate design that included temperature as a continuous factor. Following \cite{lukemire2019}, here we considered the model $Y \sim \operatorname{Bernoulli}(\mu)$ with 
{\small
\begin{eqnarray*}
\operatorname{logit}(\mu)=\beta_0+\beta_1 Algae + \beta_2 Scavenger +\beta_3 Resin +\beta_4 Compatibilizer +\beta_5 Temperature. 
\end{eqnarray*}
}
The resulting design outperformed discretized approaches, showing higher D-efficiency and robust sensitivity profiles. In order to compare two designs, we consider the relative $D$-efficiency of ${\xi}$ is with respect to the optimal design, ${\xi}^*$, defined as
\begin{eqnarray*}
\left(\frac{\operatorname{det}\left(I_{\xi}\right)}{\operatorname{det}\left(I_{\xi^*}\right)}\right)^{1 / k}     
\end{eqnarray*}
for a GLM with $k$ parameters in the linear predictor. Here $\psi$ is a design with support points at $\boldsymbol{x_1}, \ldots, \boldsymbol{x_m}$ for which there are $n_i$ replicates at each support point ${x}_i$, and ${I}$ is the Fisher information matrix calculated as
\begin{eqnarray}\label{eq.5.1.FI}
{I}_\xi=\sum_{i=1}^m n_i \Upsilon\left(\eta_i\right) \boldsymbol{x}_i \boldsymbol{x}_i^T,
\end{eqnarray}
where $\Upsilon\left(\eta_i\right)=\frac{\left(d \mu_i / d \eta_i\right)^2}{\mu_i\left(1-\mu_i\right)}$. Our algorithm aims to identify the optimal allocation of the resources by identifying the proportions 
$$p_i = \frac{n_i}{\sum_{i=1}^m n_i}.$$ 
This design in reported in Table~\ref{tab:odor} with $m=8$. As indicated by the sensitivity plot (following the framework of \citet{fedorov1972}, as discussed in Section~\ref{Sec2.4:eq} but not reported here), although our $p$-PSO design is not exactly the optimal design, which is expected due to the construction of the algorithm, it is nevertheless highly efficient. In fact, here the D-efficiency of the design generated by our proposed $p$-PSO algorithm, relative to the design produced by the $d$-QPSO algorithm \cite{lukemire2019}, is 
$$(0.8243)^{\frac{1}{6}}=0.9683.$$ 
It is interesting to note the optimal $d$-QPSO design has $m=14$ support points. From a practical standpoint, practitioners may therefore prefer this 97\% efficient design over the optimal design due to reduced number of support points, which can be useful if there are factors where it is difficult/costly to change settings.

\begin{table}[h]
\centering
\caption{The $p$-PSO design for the odor removal experiment}
{\small
\begin{tabular}{crrrrrr}
\hline
Support Point & Algae & Scav. & Resin & Comp. & Temp. & $p_i$ (\%) \\
\hline
1&	$-1$&	$ 1$&	$ 1$&	$-1$&	16.73&	14.03\\
2&	$-1$&	$ 1$&	$-1$&	$ 1$&	23.31&	 8.48\\
3&	$ 1$&	$ 1$&	$ 1$&	$-1$&	 5.00&	12.85\\
4&	$-1$&	$-1$&	$ 1$&	$ 1$&	35.00&	10.68\\
5&	$-1$&	$ 1$&	$-1$&	$ 1$&	 5.00&	14.19\\
6&	$ 1$&	$-1$&	$ 1$&	$ 1$&	 5.00&	14.52\\
7&	$-1$&	$-1$&	$-1$&	$-1$&	20.11&  15.71\\
8&	$-1$&	$ 1$&	$ 1$&	$ 1$&	35.00&   9.54\\ 
\hline
\end{tabular}
}
\label{tab:odor}
\end{table}

\begin{figure}[ht]
    \centering
    \includegraphics[width=0.5\textwidth]{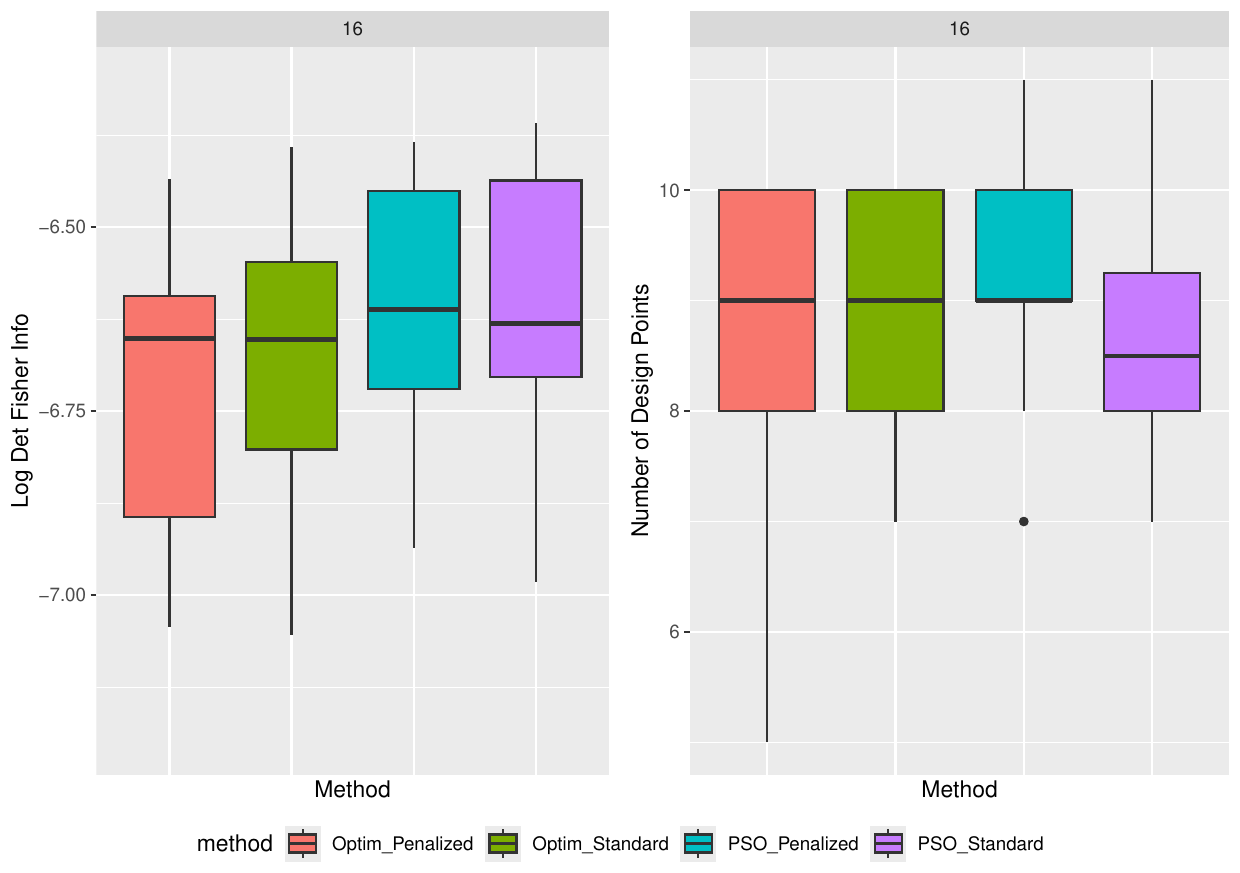}
    \caption{Comparison of penalized PSO ($p$-PSO) with standard PSO and optimization-based methods for the odor removal experiment.}
    \label{fig:odor}
\end{figure}

Moreover, we evaluated the performance of the proposed algorithm through simulation studies. Figure~\ref{fig:odor} summarizes the results based on 50 repeated runs of the algorithms. The left panel displays the log-determinant of the Fisher information matrices, while the right panel presents the corresponding number of support points. Overall, the results demonstrate that the performance of the $p$-PSO algorithm is highly satisfactory.

\subsection{Electrostatic Discharge Experiment}
\begin{figure}[h]
    \centering
    \includegraphics[width=0.5\textwidth]{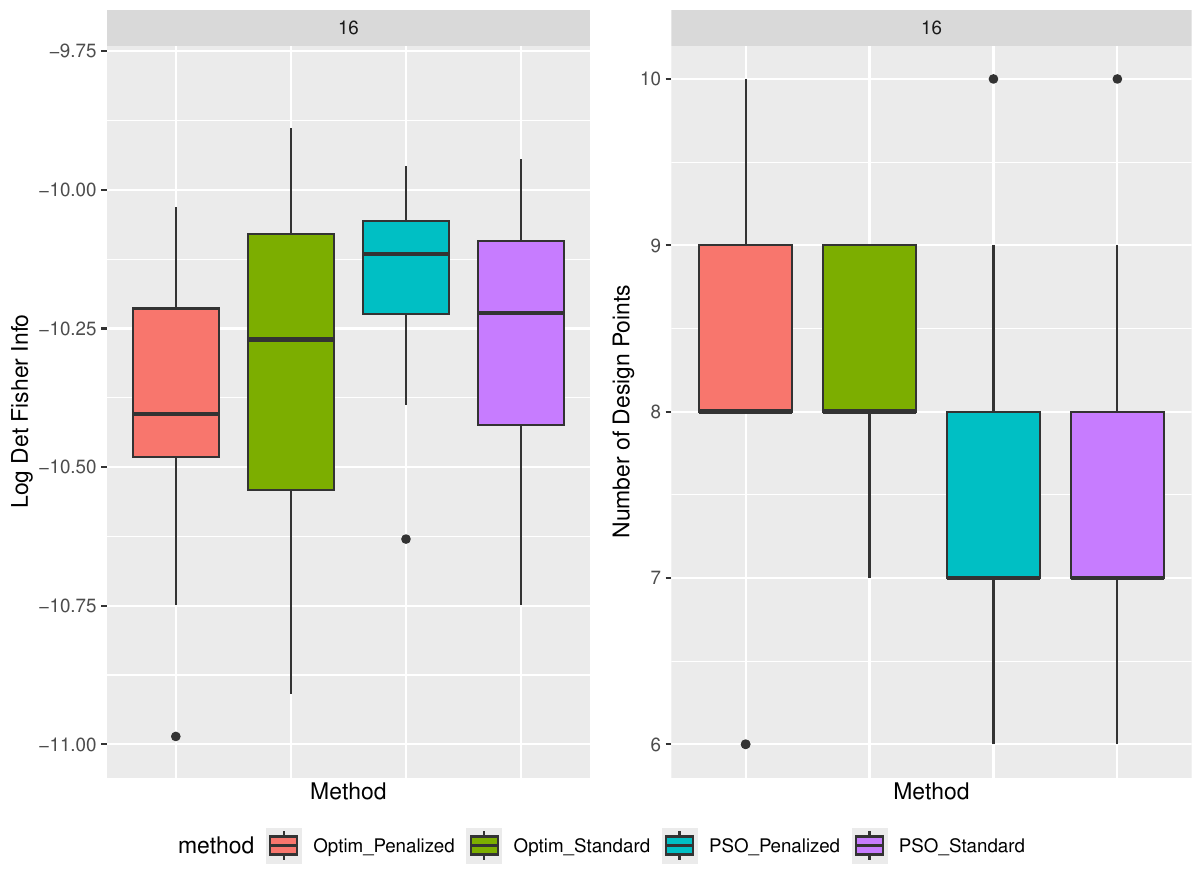}
    \caption{Comparison of penalized PSO ($p$-PSO) with standard PSO and optimization-based methods for the electrostatic discharge experiment.}
    \label{fig:car}
\end{figure}

Whitman et al. (2006) studied semiconductor failure under electrostatic discharge with discrete factors (wafer lot positions, ESD, pulse type) and a continuous voltage factor. The original design discretized voltage into five levels. Our $p$-PSO design treated voltage as continuous, identifying optimal settings that improved D-efficiency by more than 200\% relative to the original factorial design. Similar to the odor removal example, here also the response is binary, indicating whether or not a chip fails in a trial. For $Y \sim \operatorname{Bernoulli}(\mu)$, with the model 
{\small
$\operatorname{logit}(\mu)=\beta_0+\beta_1 \operatorname{Lot} \mathrm{~A}+\beta_2 \operatorname{Lot} \mathrm{~B}+ \beta_3 \mathrm{ESD}+\beta_4 Pulse +\beta_5 Voltage +\beta_{34}(\mathrm{ESD} \times Pulse ).$   
}
Following \cite{lukemire2019}, we considered the parameter vector $\beta = (-7.5$, $1.5$, $-0.2$,$-0.15$, $0.25$, $0.35$, $0.4)^T$ for the computation of the locally optimal design. As before, although the $p$-PSO design is not exactly optimal, it is highly efficient. The D-efficiency of the $p$-PSO design relative to the $d$-QPSO design is $(0.8261)^{1/7} = 0.9731$. Here also the $p$-PSO design has fewer support points ($m=9$ compared to $m=13$ for $d$-QPSO design) and hence is more desirable. The simulation results, summarized in Figure~\ref{fig:car}, exhibit a similar pattern to that observed in the odor experiment.

\begin{table}[h]
\centering
\caption{The $p$-PSO design for the electrostatic discharge experiment}
{\small
\begin{tabular}{crrrrrr}
\hline
Support Point & A & B & ESD & Pulse & Volt & $p_i$ (\%) \\
\hline
1& $-1$& $ 1$& $-1$& $ 1$& 25.00&  9.27\\
2& $ 1$& $ 1$& $ 1$& $-1$& 25.00& 14.29\\
3& $-1$& $ 1$& $ 1$& $ 1$& 25.00&  9.74\\
4& $-1$& $ 1$& $-1$& $-1$& 25.00&  9.61\\
5& $-1$& $-1$& $ 1$& $ 1$& 25.00&  9.27\\
6& $-1$& $ 1$& $ 1$& $-1$& 33.35& 14.29\\
7& $-1$& $-1$& $-1$& $ 1$& 25.00&  9.74\\
8& $-1$& $-1$& $ 1$& $-1$& 25.00& 14.29\\
9& $-1$& $-1$& $-1$& $-1$& 25.00&  9.51\\ 
\hline
\end{tabular}
}
\label{tab:elec}
\end{table}

\subsection{Car Refueling Experiment}

\begin{figure}[ht]
    \centering
    \includegraphics[width=0.5\textwidth]{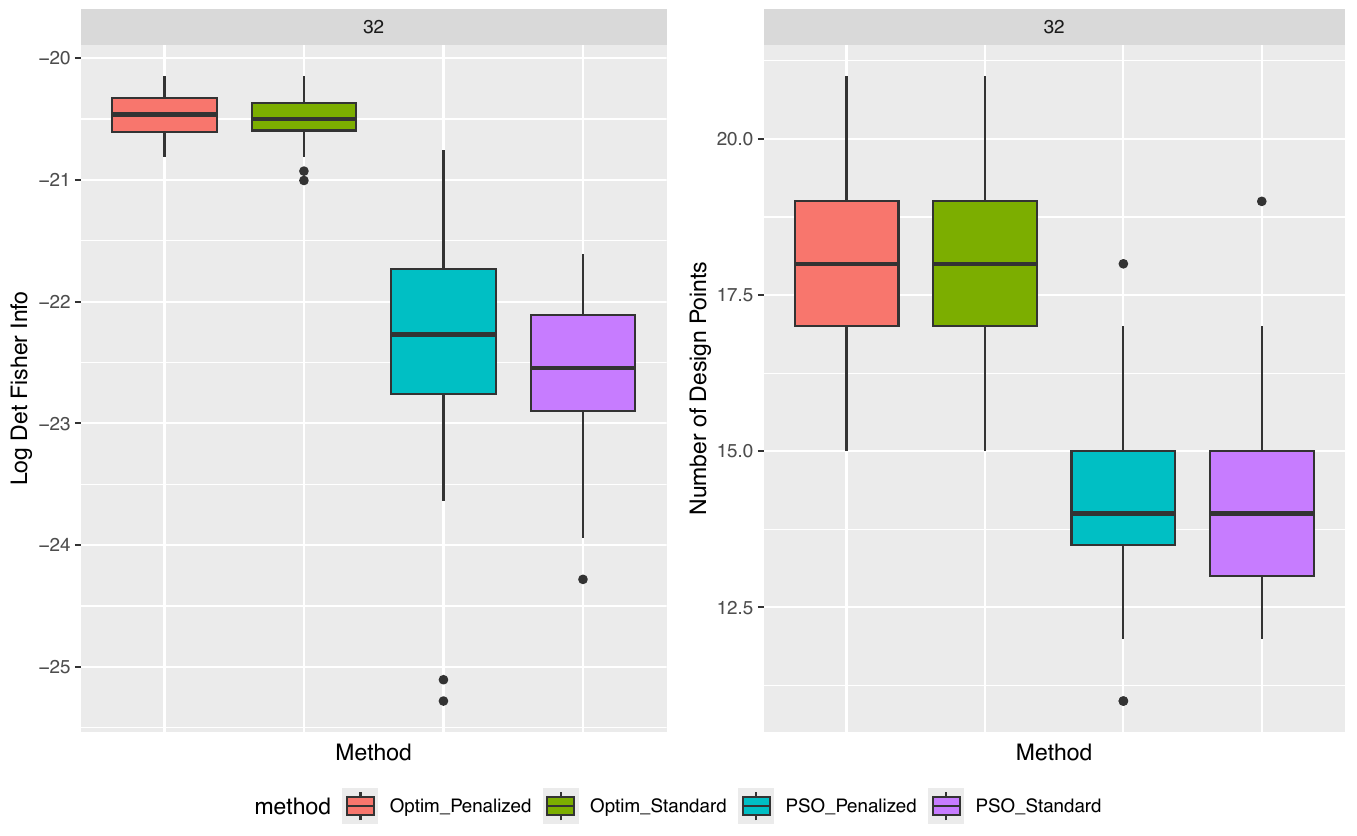}
    \caption{Comparison of penalized PSO ($p$-PSO) with standard PSO and optimization-based methods for the Car Refueling Experiment.}
    \label{fig:carplot}
\end{figure}

As described in Section~\ref{Sec4:pPSO}, \citet{grimshaw2001} conducted an experiment to evaluate a vision-based car refueling system. The aim was to determine whether a computer-controlled nozzle could be successfully inserted into the fuel pipe, producing a binary response. The locally D-optimal approximate design generated by the $p$-PSO algorithm for the standard main-effects logit model, with nominal parameter values $\beta = (3, 0.5, 0.75, 1.25, 0.8,$ $0.5, 0.8, -0.4, -1.00, 2.65, 0.65)^T$, is reported in Table~\ref{tab:car}. As in the previous example, the $p$-PSO design is highly efficient, though not strictly optimal. The D-efficiency of the $p$-PSO design $(m=11)$ relative to the $d$-QPSO design $(m=12)$ is $(0.56252)^{1/11} = 0.949$. In addition to having fewer support points and high efficiency, an attractive feature of the $p$-PSO design is that it is almost uniform, with $p_i \approx \tfrac{1}{11}$ for all $i$. Simulation studies, summarized in Figure~\ref{fig:carplot}, exhibit a similarly satisfactory performance to that observed previously.

\begin{table}[h]
\centering
\caption{The $p$-PSO design for the car refueling experiment}
\resizebox{\columnwidth}{!}{
\begin{tabular}{crrrrrrrrrrcc}
\hline
Support & Ring & & & & Lighting & & Y-axis  &Car & Ring & Threshold & \\
Point &  type & Lighting & Sharpen & Smooth & angle & Z-axis & skew & dist. & thick & stepsize & $p_i$ (\%) \\
\hline
 1 & $ -1$ &$ -1$ &$-1$ &$ -1$ &$ 54.64074$ &$ 30.00000$ &$ 10.00000$ &$ 48.00000$ &$ 0.125$ &$5.00000$ & $9.10$ \\
 2 & $-1$ &$1$ &$-1$ &$-1$ &$ 50.00000$ &$30.00000$ &$10.00000$ &$48.00000$ &$0.125$ &$ 5.00000$ & $9.10$ \\
3 & $-1$ &$-1$ &$-1$ &$-1$ &$ 50.00000$ &$30.00000$ &$0.00000$ &$48.00000$ &$0.125$ &$ 5.00000$ & $9.10$ \\
4 & $-1$ &$-1$ &$-1$ &$-1$ &$ 50.00000$ &$30.00000$ &$10.00000$ &$48.00000$ &$0.125$ &$ 8.56980$ & $9.10$ \\
5 & $-1$ &$-1$ &$-1$ &$-1$ &$ 50.00000$ &$30.00000$ &$10.00000$ &$48.00000$ &$0.425$ &$ 5.00000$ & $9.10$ \\
6 & $1$ &$-1$ &$-1$ &$-1$ &$ 50.00000$ &$30.00000$ &$10.00000$ &$48.00000$ &$0.125$ &$ 5.00000$ & $9.10$ \\
7 & $-1$ &$-1$ &$1$ &$-1$ &$ 50.00000$ &$30.00000$ &$10.00000$ &$48.00000$ &$0.125$ &$ 5.00000$ & $9.10$ \\
8 & $-1$ &$-1$ &$-1$ &$-1$ &$ 50.00000$ &$30.00000$ &$10.00000$ &$48.00000$ &$0.125$ &$ 5.00000$ & $9.10$ \\
9 & $-1$ &$-1$ &$-1$ &$1$ &$ 50.00000$ &$30.00000$ &$10.00000$ &$48.00000$ &$0.125$ &$ 5.00000$ & $9.10$ \\
10 & $-1$ &$-1$ &$-1$ &$-1$ &$ 50.00000$ &$32.90046$ &$10.00000$ &$48.00000$ &$0.125$ &$ 5.00000$ & $9.00$ \\
11 & $-1$ &$-1$ &$-1$ &$-1$ &$ 50.00000$ &$30.00000$ &$10.00000$ &$45.67964$ &$0.125$ &$ 5.00000$ & $9.10$ \\
\hline
\end{tabular}}
\label{tab:car}
\end{table}

\section{Concluding Comments and Future Research}

Metaheuristic algorithms such as Genetic Algorithms (GA), Differential Evolution (DE), and Simulated Annealing (SA) have all been successfully employed for constructing optimal experimental designs. Our proposed penalized PSO ($p$-PSO) contributes to this growing body of work by offering a simple yet effective way to adapt standard PSO for mixed discrete–continuous problems. 

Compared to $d$-QPSO, $p$-PSO offers several important advantages. By incorporating penalty terms directly into the objective function, the method provides enhanced constraint handling and ensures that infeasible solutions are systematically avoided. This leads to improved convergence stability, particularly in experiments with both discrete and continuous factors, where traditional swarm optimization methods often struggle. The framework is flexible and can be applied to both exact and approximate designs, and it naturally extends to pseudo-Bayesian settings. Although $p$-PSO is computationally somewhat more demanding than standard PSO, the additional cost is modest and outweighed by the robustness it provides. In our case studies, $p$-PSO consistently delivered designs with higher efficiencies than competing approaches.

The key idea underlying $p$-PSO is simple yet powerful: by penalizing the objective function according to the distance from valid discrete levels, the mixed discrete–continuous design problem is effectively transformed into a continuous one. This allows the use of standard, off-the-shelf PSO implementations without the need for specialized update rules for categorical variables. The simplicity of this approach suggests that it may also be useful for optimization problems outside the domain of experimental design.

There remain many promising directions for future work. One natural extension is to explore adaptive penalty schemes that adjust the strength of penalization dynamically during the search process. Beyond the $D$-optimality criterion, $p$-PSO could be adapted to other design criteria such as $A$- and $E$-optimality, thereby broadening its applicability. Incorporating interactions among factors represents another important area, since more realistic models may include complex relationships between variables. Efficiently handling discrete factors with more than two levels is also of interest, and alternative penalty formulations could be explored to facilitate smooth transitions between levels, for example encouraging direct movement from low to high settings without passing through intermediate values. Finally, future research may test other metaheuristic algorithms, such as simulated annealing, within the penalized framework, as well as extend the method to high-dimensional design problems with many interacting factors.

In summary, we have proposed a penalized PSO algorithm for finding D-optimal designs for GLMs with mixed factors. The method is easy to implement, general in scope, and has the potential to contribute beyond the design of experiments to the broader field of constrained optimization.


\section*{Acknowledgements}
This research was partially supported by the NSF Grant DMS-2311186. The authors would like to thank Dr. Joshua Lukemire for several fruitful discussions and for generously sharing his time and ideas, which greatly benefited this work. We are also grateful to Dr. Vinod Gupta for his encouragement throughout the course of this research.

\section*{Conflict of interest}
The authors do not have any financial or non-financial conflict of interest to declare for the research work included in this article.

\bibliographystyle{chicago}  
\bibliography{ref}


\appendix
\section*{Appendix}

\begin{algorithm}[H]
\caption{p-PSO for D-Optimal Design}
\begin{algorithmic}[1]
\Require GLM with parameter vector $\boldsymbol{\beta}$; design space $\mathcal{X}$
         with $q$ discrete and $r$ continuous factors; swarm size $N$;
         number of support points $m$; iterations $T$;
         inertia weight $w$; penalty parameter $\lambda$
\Ensure  Best feasible design $\texttt{gbest}$

\vspace{4pt}
\State \textbf{// Step 1: Initialization}
\For{each particle $i = 1, 2, \ldots, N$}
    \State  Initialize position $x_i$ randomly in search space
    \State Initialize velocity $v_i$ randomly
    \State Project $x_i$ to nearest valid design $\tilde{x}_i$:
           replace each discrete factor by its nearest valid level
    \State Compute penalized fitness:
           $F(x_i) = \Phi(\tilde{x}_i) - \lambda P(x_i)$
    \State Set $\texttt{pbest}_i \leftarrow x_i$
\EndFor
\State Set $\texttt{gbest} \leftarrow \arg\max_i\, F(x_i)$

\vspace{4pt}
\State \textbf{// Step 2: Main loop}
\For{$t = 1, 2, \ldots, T$}
    \For{each particle $i$}
        \State Draw $r_1, r_2 \sim U(0,2)$ independently at each iteration
        \State Update velocity:
               $v_i \leftarrow w\,v_i + r_1(\texttt{pbest}_i - x_i)
               + r_2(\texttt{gbest} - x_i)$
        \State Update position: $x_i \leftarrow x_i + v_i$
        \State Project to nearest valid design $\tilde{x}_i$
        \State Compute penalized fitness:
               $F(x_i) = \Phi(\tilde{x}_i) - \lambda P(x_i)$
        \If{$F(x_i) > F(\texttt{pbest}_i)$}
            \State $\texttt{pbest}_i \leftarrow x_i$
        \EndIf
        \If{$F(\texttt{pbest}_i) > F(\texttt{gbest})$}
            \State $\texttt{gbest} \leftarrow \texttt{pbest}_i$
        \EndIf
    \EndFor
\EndFor

\vspace{4pt}
\State \textbf{// Step 3: Return best feasible solution}
\State Snap all discrete factor levels in \texttt{gbest} to nearest valid levels; normalize weights to sum to 1
\State \Return $\texttt{gbest}$
\end{algorithmic}
\end{algorithm}

\end{document}